\documentclass[prb,twocolumn,floatfix,showpacs,superscriptaddress]{revtex4}
\usepackage{graphicx}
\usepackage{amsmath}

\begin{document}

\title{Implementing the Keldysh formalism into the {\it ab initio} 
Gaussian Embedded Cluster Method for the calculation of quantum transport}
\author{E. Louis}
\affiliation{Departamento de F\'{\i}sica Aplicada, Universidad de
Alicante, San Vicente del Raspeig, Alicante 03690, Spain.} 
\affiliation{Unidad Asociada del Consejo Superior de Investigaciones
Cient\'{\i}ficas, Universidad de
Alicante, San Vicente del Raspeig, Alicante 03690, Spain.} 
\author{J. A. Verg\'es}
\affiliation{Departamento de Teor\'{\i}a de la Materia Condensada, Instituto
de Ciencia de Materiales de Madrid (CSIC), Cantoblanco, Madrid 28049, Spain.}
\author{J.~J.~Palacios}
\affiliation{Departamento de F\'{\i}sica Aplicada, Universidad de
Alicante, San Vicente del Raspeig, Alicante 03690, Spain.}
\affiliation{Unidad Asociada del Consejo Superior de Investigaciones
Cient\'{\i}ficas, Universidad de
Alicante, San Vicente del Raspeig, Alicante 03690, Spain.} 
\author{A. J. P\'erez-Jim\'enez}
\affiliation{Departamento de Qu\'{\i}mica-F\'{\i}sica, Universidad de
Alicante, San Vicente del Raspeig, Alicante 03690, Spain.}
\author{E. SanFabi\'an}
\affiliation{Unidad Asociada del Consejo Superior de Investigaciones
Cient\'{\i}ficas, Universidad de
Alicante, San Vicente del Raspeig, Alicante 03690, Spain.} 
\affiliation{Departamento de Qu\'{\i}mica-F\'{\i}sica, Universidad de
Alicante, San Vicente del Raspeig, Alicante 03690, Spain.}
\begin{abstract}
We discuss the key steps that have to be followed to calculate
quantum transport out of equilibrium by means of the {\it ab initio} 
Gaussian Embedded Cluster Method recently developed by the authors.
Our main aim is to emphasize through several examples that, 
if a sufficiently large 
portion of the electrodes is included in the {\it ab initio} calculation,
which does also incorporate an electrochemical potential difference   
$\mu_L-\mu_R=eV$,
there is no need to impose an electrostatic potential $V$ drop
accross the system.
\end{abstract}
\maketitle

The effort devoted to investigate electronic transport through 
molecular bridges and metallic nanocontacts has sharply
increased in recent years\cite{Joachim:nature:01,Nitzan:arpc:01,Agrait:pr:02}. 
In particular, a variety of {\it ab initio}
methods are being developed aiming to catch the essentials of these 
systems\cite{Lang:prb:95,Mujica:jcp:00,Taylor:prb:01:b,
Xue:condmat:01,Damle:prb:01,Brandbyge:prb:02,Damle:condmat:02}. 
These methods share most of the ingredients 
[Landauer-Keldysh formalism for
transport and density functional (DF) theory], but differ in 
their numerical implementation, which translates into their 
computational efficiency and predictibility 
power. 
A critical problem not yet fully solved is the way these
methods actually deal with out-of-equilibrium 
transport\cite{Keldysh:jetp:65,Meir:prl:92,Hershfield:prl:91}. Most
of the approaches that are being  proposed impose from
the outset an external and uniform electrostatic  potential
drop across the molecule or nanocontact \cite{Taylor:prb:01:b,
Damle:prb:01,Brandbyge:prb:02}. The latter is justified
when considering planar electrodes, which 
in most cases do not correspond to  experimental 
electrode geometries. This choice simplifies the numerical implementation, but
can condition the outcome of the calculation.
In this work we discuss  how the Keldysh formalism can
be implemented into {\em ab initio} transport schemes and, in particular, into
the Gaussian Embedded Cluster Method (GECM) recently
developed by the authors\cite{Palacios:prb:01,Palacios:prb:02}. 
Conclusive numerical evidence is
presented which shows that, if a sufficiently large portion
of the electrodes is incorporated into the {\it ab initio}
calculations, 
there is no need to add an external 
electrostatic potential to the self-consistent potential.
This provides a natural, consistent, way of solving the electrostatic
problem for generic electrode geometries.  

In a single-particle description, inherent to DF calculations,
one obtains the current $I$ through the
molecule or nanocontact at non-zero temperature and 
finite bias voltage $V$ by integrating the transmission probability:
\begin{equation}
I=\frac{e}{h}\int_{-\infty}^{\infty} T(E,V)
 [f(E-\mu_L)-f(E+\mu_R)] {\rm d}E,
\label{I}
\end{equation}
where $f(E-\mu_{\rm L,R})$ is the Fermi distribution for the left and right
electrodes whose electrochemical potentials are denoted by
$\mu_{\rm L}$ and $\mu_{\rm R}$, respectively, and $\mu_{\rm L}-\mu_{\rm R}=eV$.
The transmission coefficient $T(E,V)$ depends 
both on temperature and the applied bias, and is  given by
\begin{equation}
T(E,V) =
{\rm Tr}[\hat\Gamma_{\rm L}\hat G^{(+)} \hat\Gamma_{\rm R}\hat G^{(-)}],
\end{equation}
where hats  denote operators. The Green function operators and the 
gamma operators depend on energy and the applied potential (both 
arguments dropped for convenience). The latter are defined as
$\hat\Gamma_{\rm L,R}=\hat\Sigma_{\rm L,R}^{(+)}-\hat\Sigma_{\rm L,R}^{(-)}$,
where $\hat\Sigma_{\rm L,R}$ denotes the self-energy operator
for the part of the right and left semi--infinite electrodes which is not
included in the {\it ab initio} calculation. The Green function operator
is in its turn defined as
\begin{equation}
\left [(E\pm i\delta)\hat I-\hat F - \hat\Sigma^{(\pm)}_L-\hat\Sigma^{(\pm)}_R
\right ] \hat G^{(\pm)}= \hat I ,
\label{green}
\end{equation}
\noindent where ${\hat I}$ is the unity operator.
On the other hand, the density matrix out of equilibrium is
obtained from\cite{error}
\begin{equation}
P_{\alpha\beta}=-\frac{i}{2\pi}\int_{-\infty}^{\infty}
{\rm d} E \sum_{\gamma \delta}\left[S^{-1}_{\alpha \gamma} 
G^{<}_{\gamma \delta} S^{-1}_{\delta \beta} \right ]
\label{eqn:nab}
\end{equation}
where $G^{<}_{\gamma \delta}$  are the matrix elements in a non-orthogonal
basis of the lesser Green function operator given by 
\begin{equation}
{\hat G}^<=i{\hat G}^{(+)}\left[f(E-\mu_L){\hat \Gamma}_{\rm L}+
f(E-\mu_R){\hat \Gamma}_R\right]{\hat G}^{(-)}.
\end{equation}
Most of the technical difficulties implicit in the evaluation of the 
above expressions have been discussed in previous 
works\cite{Taylor:prb:01:b,Damle:prb:01,Palacios:prb:02,Brandbyge:prb:02}. 
There is, however, an important conceptual 
issue that, in our opinion, remains unsolved and worth clarifying. 
In the presence of a voltage difference between electrodes
one might be tempted to impose from the outset an
external electrostatic potential drop across the molecule or nanocontact.
As the potential profile 
depends strongly on the electrode geometry, one would have to
solve Poisson equation with the boundary conditions appropriate for such
geometry. A way around this problem is to assume a simple form for the 
electrodes\cite{Taylor:prb:01:b,Damle:prb:01,Brandbyge:prb:02}.
However,  adding an external potential through the homogeneous
solution of the Poisson equation is unnecessary 
if a significant part of the metallic electrodes has already
been included in the initial cluster and an {\em electrochemical 
potential difference} is maintained. The difference between left and right 
electrochemical potentials charges one electrode and discharges
the other one, effectively creating an electrostatic
potential drop across the constriction\cite{Landauer:ibm:57}. 
In general, once self-consistency 
is attained, an electrostatic potential difference $V$ between 
atoms one or two layers inside opposite electrodes develops while
they remain neutral. On the other hand, the potential difference between
atoms on the surface of opposite
electrodes is smaller than $V$ since they carry the charges. In
summary, we assume all the charges responsible for the electrostatic potential
across the constriction to be part of our cluster instead of imposing a usually
unknown external electrostatic potential.  For a
large enough number of atoms in the electrodes this should be essentially
correct. 
\begin{figure}[t]
\includegraphics[width=2.5in,height=3.5in]{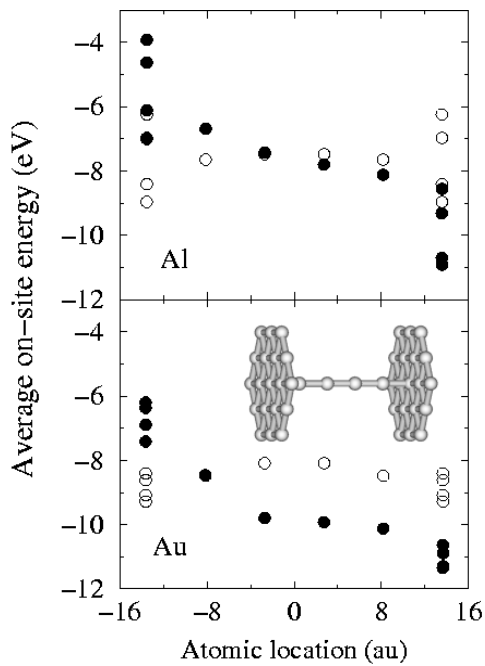}
\caption{Average onsite energies of the atomic orbitals versus atomic position
in  $M19-M4-M19$ nanocontacts (two 19 atom (111)-planes plus a four atom 
chain, see inset). The  results correspond to $V$=0 (empty
circles) and 5 V (filled circles) and $M\equiv$Al (a) and Au (b).  
\label{onsite-111}}
\end{figure}
\begin{figure}[t]
\includegraphics[width=2.5in,height=3.0in]{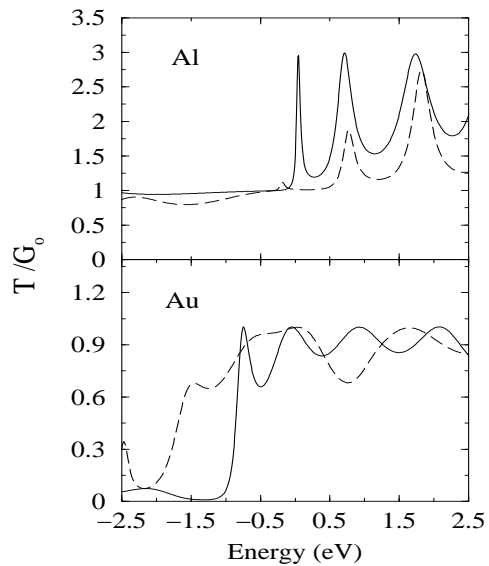}
\caption{Transmission versus energy for $V$=0 (continuous line)
and 5 V (dashed line) in the $M19-M4-M19$ nanocontacts of Figure 1, with 
$M\equiv$Al (a) and Au (b), are shown. 
\label{trans-111}}
\end{figure}

The way we have actually implemented this self-contained, 
out-of-equlibrium scheme into the GECM initially
proposed in Ref.~\onlinecite{Palacios:prb:01} and later fully developed in
Ref.~\onlinecite{Palacios:prb:02} is hereafter described.
A first step consists in carrying out a DF calculation by means
of the GAUSSIAN98 code\cite{Gaussian:98} of the region that contains
the molecule or the set of atoms that form the contact between
the electrodes plus {\it a significant part of the electrodes}. Once 
a reasonable accuracy has been achieved, the Green function for 
the infinite system is calculated and the density matrix given
by Eq.~\ref{eqn:nab} used in the subsequent DF process. The infinite electrode
selfenergies are calculated by means of the Bethe lattice
approximation and a minimal basis set with pseudopotentials was
used in the calculations [see Ref.~\onlinecite{Palacios:prb:02} for details]. 
In previous works, the DF
calculations were carried out by means of the Becke
three-parameter hybrid functional plus the correlation functional 
of Lee, Yang and Parr (B3LYP)\cite{Becke:jcp:93}. A difficulty specific
to the out-of-equilibrium case comes from the fact that the density matrix
in Eq.~\ref{eqn:nab} is complex. This can be handled in two ways: i) treating
exchange within DF, instead of carrying out a full Hartree-Fock
calculation (this can be easily done by using BLYP instead of B3LYP),  and
ii) incorporating
the complex density matrix into the GAUSSIAN98 code. In the present case
we have chosen the former 
as the second procedure would have required  major changes
in the standard code\cite{link}.
In order to get reasonably reliable results for finite
bias, a root mean square error in the density matrix better
than $10^{-4}$ was mandatory. 
\begin{figure}
\includegraphics[width=2.5in,height=3.5in]{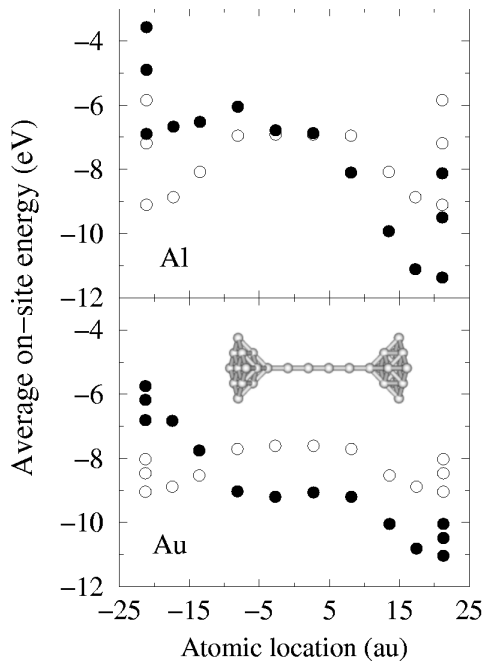}
\caption{Same as Figure 1 for a (100) pyramidal arrangement (two pyramids
containing 14 atoms each plus a four-atom chain; see inset) 
referred to as $M14-M4-M14$ (see text).
\label{onsite-100}}
\end{figure}

Calculations have been carried out  for gold and 
aluminum nanocontacts similar to those of the insets in Figures 1 and 3. 
Two types of fcc clusters have been considered: (a) two (111) planes 
containing 19 atoms each plus a  four atom chain labeled $M19-M4-M19$
(a capacitor was described by taking out the four atom chain leaving
a wide gap between the planes); (b) two (100)-oriented 
pyramidal clusters plus a 4 atom chain
labeled $M14-M4-M14$ [(100)-oriented pyramidal nanocontacts 
as $M5-M5$, $M14-M14$ 
and $M30-M30$ have also been considered]. Interatomic bulk distances have been
taken for the whole cluster (2.86 $\AA$ for Al and 2.88 $\AA$ for  Au).
All results correspond to zero
temperature and an external bias voltage in the range 0-5 V. The structural
stability of the system models is an important issue when one is interested
in the interpretation of experimental results. This is not the case here.

Figure 1 shows the average on-site energies ($5d 6s 6p$ for Au and
$3s 3p$ orbitals for Al) on all atoms of the 
nanocontacts $M19-M4-M19$ for zero  and  5 V bias.
This magnitude reflects only 
the electrostatic potential on each atom, not telling
us anything about the chemical potential (charge) on them.
The on-site energies in the planes are less dispersed in the case of 
gold than in
aluminum. This is probably due to the simple $s$-character of the 
wavefunctions at the Fermi level that gold has (as opposed to the
$sp$ character in aluminum). At 5 V bias, the major drop in the potential
occurs at the chain/plane contacts. While 
the total potential drop in gold (aluminum) is 4.44 eV (4.6 eV)
the drop between the first and the last chain atoms is
only 1.67 eV (1.43 eV). In the case of zero bias,
the results are symmetric with respect to the geometric center of 
symmetry, as expected. Instead, a similar symmetry is absent for 5 V. 
Namely, whereas the potential drop between the left electrode
and the first atom in the chain is 1.92 eV  (1.69 eV) for gold (aluminum)
nanocontacts, it is  only of 0.85 eV (1.48 eV) between the 
chain end and the right electrode.
This feature seems common to all previously reported results
and reflects not only bulk band structure features, but also 
details of the nanocontact geometry (see below).
The transmission for $V=0$ and 5 V is shown in Figure 2. There are noticeable
differences both in gold and aluminum. In the latter the peak structure is 
significantly changed, whereas in gold the gap below -0.5 eV is partially
filled. It is worth noting  that although the total
current calculated  by integrating in a window [-$V$/2, $V$/2] 
the transmission for $V$=0 
does not differ much from that obtained with the full
non-equilibrium approach (more important changes are expected
in complex molecules), the differential conductance
is significantly different. 
\begin{figure}
\includegraphics[width=2.5in,height=3.0in]{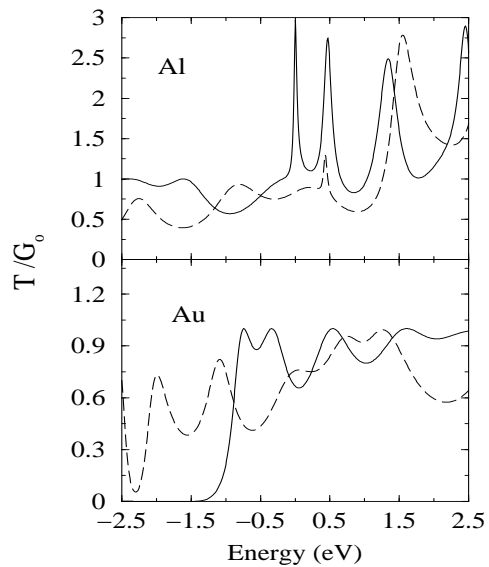}
\caption{Transmission versus energy for $V$=0 (continuous line)
and 5 V (dashed line) in the  $M14-M4-M14$ nanocontacts of Figure 3, 
with $M$= Al (a) and Au (b). 
\label{trans-100}}
\end{figure}

The average on-site energies on all atoms of the 
nanocontacts $M14-M4-M14$ for zero and  5 V bias,
are depicted in Figure 3.  As above, the on-site energies
in the outer  planes are less dispersed in the case of 
gold than in aluminum.  Now the potential drop along
the chain is significantly different for the two metals: while in the case of
gold the total potential drop  is 4.29 eV and 
the drop between the first and the last chain atoms is
only 0.162 eV, in  aluminum these numbers change to 4.56 and 2.043, 
respectively. These results support our claim in the sense that what
matters are not only the bulk features of the metals but also
the nanocontact geometry. The asymmetry in the potential drop remarked
above is also noted. Again, the differences in the transmission
between  zero and 5 V bias are  significant. The peak structure in
aluminum is noticeably reduced and the gap in gold filled over a range
larger than 1 eV.
\begin{table}
\caption{
Fittings of the numerical results for the variation with applied
electrostatic potential $V$ of the difference between the average on-site 
energies of the atomic orbitals at the outer planes
in  clusters similar to those of Figures 1 and 2 (see text)
and of current through them,
with functions $\Delta E=aV^b$ and $I/G_0=cV^d$, respectively ($\Delta E$,
$I/G_0$ and $V$ given in eV, $G_0$ being the quantum of conductance $2e^2/h$).}
\label{Table}
\begin{tabular}{|l||cc|cc|}
\colrule
 cluster     &    a   &   b   &   c   &   d   \\
\colrule
\colrule
 $Al5-Al5$  & 0.820  & 0.979 & 0.810 & 1.039 \\
 $Al14-Al14$  & 0.882  & 0.992 & 0.807 &  1.08 \\
 $Al30-Al30$  & 0.942  & 0.984  & 0.635 & 1.20  \\
 $Al14-Al4-Al14$  & 0.900  & 1.008 & 1.196 &  0.865 \\
 $Al19-Al4-Al19$  & 0.919  & 1.001 & 1.315 & 0.925  \\
 $Al19$-vacumm-$Al19$  & 0.953  & 1.001  & -  & - \\
\colrule
 $Au5-Au5$  & 0.803  & 0.935  & 0.521  & 1.068  \\
 $Au14-Au14$  & 0.840  &  0.993  &  0.668 & 1.067  \\
 $Au14-Au4-Au14$  & 0.882  & 0.971 & 0.805 &  0.950 \\
 $Au19-Au4-Au19$  & 0.882  & 1.003 & 0.886  & 0.91  \\
 $Au19$-vacumm-$Au19$  & 0.928  & 1.001  & -  & - \\
\colrule
\end{tabular}
\end{table}

In Table  I we report the fittings of the numerical results for the 
difference between the average
energies of all atoms in the left and right planes in 
$M19-M4-M19$ nanocontacts, and the  two outer planes
in the pyramidal nanocontacts, versus the applied external potential.
The fittings were done with  $aV^b$, instead of assuming
a linear relationship from the start. All  regression coefficients 
were higher than 0.999. Although  $a$ and  $b$ 
are always rather close to 1, some significant features are worth of comment. 
There is a steady increase of $a$ and $b$ as 
the size of the outer planes increases. The values closest to one are found for 
the capacitors (the nanocontacts with a $\approx 14 \AA$ gap). 
In the latter case 
the coefficient $a$ is only 7\% (5\%)  smaller than 1 for gold (aluminum) 
nanocontacts, while the exponent 
cannot in practice be differentiated from one. 
The small deviations with respect 
to the "ideal" behavior are  due to the finite charge that those planes
carry and/or to a yet insufficiently large
electrode. As expected, in pyramidal nanocontacts the results steadily 
approach  the ideal 
behavior as the size of the cluster increases. These results 
constitute the main message of the present work. 
The results for the current versus voltage were also 
fitted with $I/G_0=bV^d$ (see Table I). In this case
the lowest regression coefficient is 0.99 (in some 
cases the results are better fitted with second order polynomia).
The significant deviations from a
linear behavior (Ohm's law) are   not  
surprising in 
systems as small as those investigated here and have been reported 
previously. The current is 
in general higher in aluminum than in gold, which is consistent with the larger 
conductance found for aluminum at the Fermi level\cite{Palacios:prb:02}.  
Deviations from 
linearity are also more important in the case of aluminum, with the
exponent increasing with  the size
of the pyramidal cluster. This is a consequence of a number of facts, 
including the increasing contributions from the $\pi$ channel discussed in 
Ref.~\onlinecite{Palacios:prb:02}. 

Summarizing, we have discussed  the practical 
procedures one should follow to implement the non-equilibrium Keldysh formalism 
into the Gaussian Embedded Cluster Method previously developed by the authors. 
In our view the most important outcome of this work is the
conclusive evidence concerning the need of incorporating a sufficiently large 
portion of the electrodes in the {\it ab initio} calculation. If this is done, 
the artificial addition of an applied external potential to the potential in 
the Schr\"odinger equation is unnecessary. 

Partial financial support by the spanish MCYT (grants BQU2001-0883,
PB96-0085, and MAT03-04450-C03) and
the Universidad de Alicante is gratefully acknowledged. 

\bibliographystyle{apsrev}
\bibliography{prb_nolequ}

\begin{thebibliography}{20}
\bibitem[{lin()}]{link}
\bibinfo{note}{Unfortunately, the complex version of the link 502 in the
  GAUSSIAN98 code, named link 503, is rather obsolete.}
\end{thebibliography}

\end{document}